# A Strategic Investment Framework for Biotechnology Markets via Dynamic Asset Allocation and Class Diversification

Abhishek Mohan[1], Agnibho Roy[2]

[1]*Texas Academy of Mathematics and Science, Denton, Texas, 76201, United States*
[2]*Coppell High School, Coppell, Texas, 75019, United States*

**Abstract:** In this paper, we propose an innovative investment framework incorporating asset allocation and class diversification oriented specifically for the biotechnology industry. With growing interests and capitalization in multiple biotech markets, investors require a more dynamic method of managing their assets within individual portfolios for optimal return efficiency. By selecting a single firm representative of identified industry trends, analyzing financial metrics relevant to the suggested approaches, and assessing financial health, we developed an adaptable investment methodology. We also performed analyses of industrial viability and investigated the implications of the selected strategies, with which we were able to optimize our framework for versatile application within specialized biotech markets.

**Focus Points and Keywords:** Financial metrics, asset allocation, class diversification, stocks, bonds, commodities, biotechnology markets, investment frameworks, and adaptable systems.

## 1. Introduction – Investment Strategies

Investors utilize investment strategies, which consist of a systematic plan of procedures and rules to adhere to, in order to guide the allocation of their assets among a diverse range of investments, allowing for capital expansion whilst minimizing financial risk within their portfolio. Strategies primarily depend on the type of investment and the profit goal of each investment, as strategists and investors seek to find the optimal trade-off between their profit objectives and portfolio safety.[1] Traditionally, company investments in other companies serve to the benefit of both ends, but one needs to be able to pare an agreement of personal interests while minimizing associated risk so that they are able to breakeven or make a significant profit.[2] The establishment of careful asset control and diversification strategies of preferred stocks, in conjunction with analysis of financial data, allows for investors to determine which selection provides the best competitive advantage to a company. These strategies, in other terms, can be viewed as a calculated protocol of investment management, which for strategists, analysts, and independent investors alike, offers an efficient method to plan for and increase probability of fiscal success.

Although dynamic markets provide different opportunities and challenges, investors are able to choose from a wide array of possible strategies, and can modify or merge possible approaches to accommodate their intended investment objectives. For example, the branch of "passive" strategies emphasizes cautious investing, only buying and selling to develop steady profits over an extended period of time. However, by doing so, investors must resist the desire to trade their assets at times of market downturn and maintain resilience during market depreciation. Such an approach requires investors to have patience and put their faith in the market to gradually reach equilibrium. Strategies may depend on market behavior and volatility, but a general attitude should exist within the investor. As mentioned before, investors can allow market forces to dictate their decisions, but can be more aggressive within the market and attempt to participate in day trading, also known as momentum trading, where the general performance of a company is analyzed and the decisions to invest in it are accordingly made. This approach increases the risk associated within the investment, as one makes bold trades and decisions, but the short-term returns may be rewarding if the proper methods are correctly implemented.

As various tactics can be incorporated into one's own investment approaches, a generalized structure of specialized techniques allows for the specification to a definite market. One such group of rapidly developing companies is the biotechnology enterprise, which harbors cutting edge research and great prospect in future financial expansion. Identifying this area of consistent growth, this paper develops a comprehensive framework for strategic biotech investments, primarily by synthesizing asset allocation and class diversification methodologies into a joint system refined for current markets, while also allowing for dynamic adaptability within the framework.

## 2. Investigation of the Current Biotechnology Industry

As defined by the Biotechnology Innovation Organization (BIO), biotechnology, also known as biotech, "harnesses cellular and biomolecular processes to develop technologies and products that help improve our lives and the health of our planet."[3] Current developments in biotech are focused on combating global diseases and pandemics, optimizing environmental systems to reduce pollution, enhancing energy production and usage, improving manufacturing methods, and many other possible applications. The industry is relatively new, first launched by the discovery of DNA and the identification of its ability to be constructed and modified to accomplish specific objectives in 1973.[4] Since then, companies and enterprises have developed revolutionary products aimed at specific internal markets, and through progressive research, strategic planning, and specialized marketing, have established critical roles in many larger areas for market capitalization.

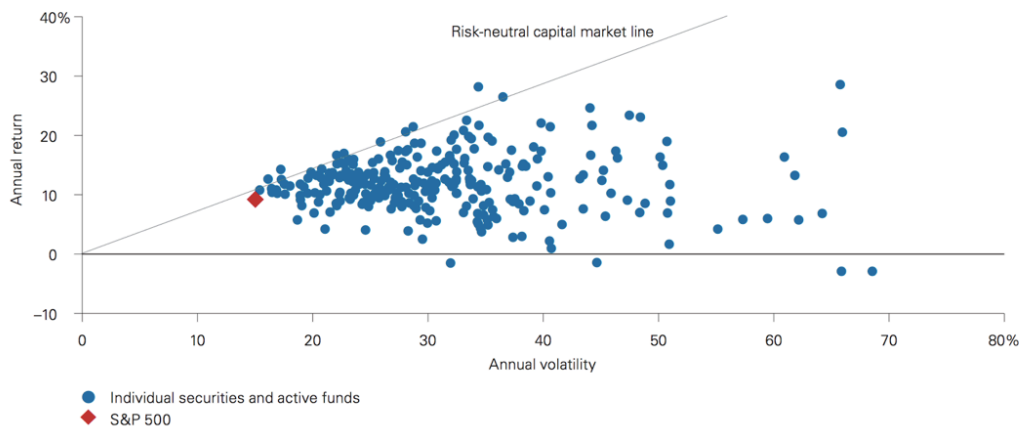

*Figure 1.* A representation of the relationship between annual volatility and annual return as presented by the Vanguard Group. Broad diversification can allow for the protection against the risks presented by owning individual securities. This further justifies the role that calculated asset allocation plays in both return and portfolio variability.[6]

Pharmaceutical companies within the biotech industry have many specialized sectors, including bioinformatics and bioremediation.[5] Such a spectrum of particular objectives allows for investors to identify what technological developments within biotech has potential for significant expansion. The primary appeal of investing in biotech companies is the prospect of creating great returns from budding technologies, providing many opportunities for strategic investing. As seen in Figure 1, trends of an increase in annual volatility suggest an increased likelihood of greater profits, supplementing investors' growing interest in biotech markets.[6] Products within this industry are capable of transitioning from minute market values to immense profits in a very short period of time. Small firms provide a chance of introducing products that could revolutionize specific areas of biotech, but the "high-risk, high-reward" sentiment should not be the only way these products are viewed through; patience and calculated speculation can provide investors with a safer, yet still effective approach for careful investing.

With the rapid growth of multiple biotech markets, investors are taking greater interest in strategic short-term and long-term investing of the industry.[7] There are a few specific considerations that must be addressed prior to developing a dynamic framework that would allow for versatile and effective asset management. Based upon Porter's 5 Forces Analysis, the first point of interest is understanding how new companies are able to enter the market.[8] Due to the biotech industry being composed of multiple smaller but developed companies, barriers to entry exist. As a result, only the most competitive and aggressive ventures are able to breakthrough, let alone succeed; such barriers, however, make it easier to identify companies with potential for rapid growth.

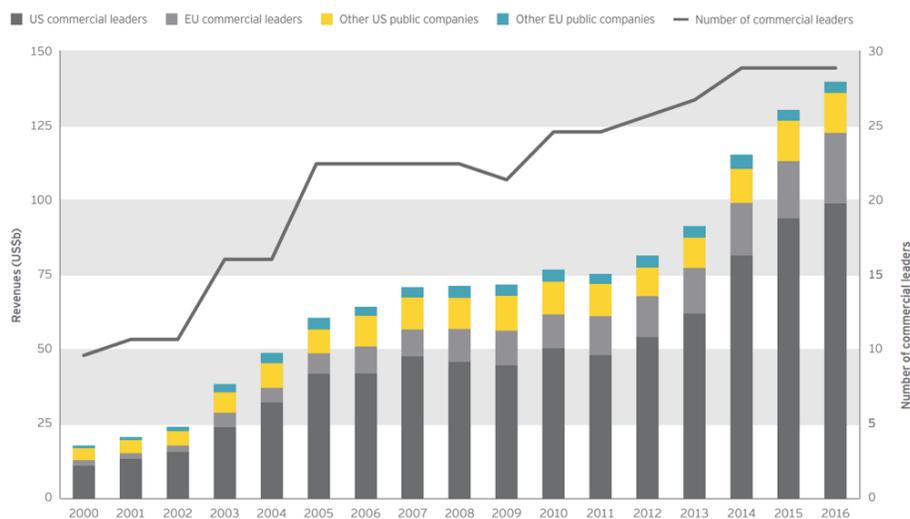

*Figure 2.* *Comparison of public company revenues and number of commercial leaders from the US and EU within the biotechnology industry, as reported by Ernst & Young (EY). Over the past 15 years, biotech markets have shown great surges in revenue and industrial growth, and are projected to continue expansion in the next decade. Widespread potential for market capitalization has attained the interest of investors who seek to develop versatile portfolios for maximum profit.*[10]

Accordingly, the wide availability of many biotech substitutes, specifically those that are easily made and readily accessible, introduce issues of price competition and should be investigated prior to any investment decision.

Specialized versus non-specialized corporations should also be differentiated prior to developing a structured framework, as customer and corporation purchasing power is dependent upon the targeted biotech market. Just as customers hold bargaining power, suppliers also hold great control of their intended target and product distribution. In many cases, private ventures license their developed technologies to prevent sharing of their intellectual property: a central source of a company's value. Lastly, the intense competition present in the biotech industry provides investors with an additional challenge, as they must carefully determine if a small venture suggesting possible expansion is more valuable than a larger, already established company.

There are a few considerations that must be made to fully understand the current state of the biotech industry. As the NASDAQ Biotech Index has recently grown despite investors' worry of volatility, many small-cap biotech firms have maintained attractive billion-dollar valuations, but not all of them exhibit accelerated growth as they once did. Many biotech firms still have multiple initial public offering (IPO) windows open due to the establishment of a few innovative technologies over the span of the past 5 years.[9] As shown in Figure 2, public company revenues in both the US and European countries have significantly grown over the past decade. This can partly be attributed to recent discoveries in CRISPR, epigenetics, and gene therapy technologies, which market analysts believe will continue to allow for expansion of biotech IPOs within the coming years.[10] There have also been multiple new revolutionary therapeutic systems, including Zydelig, CAR-T, Solvavdi, and anti-PCSK9s, which have introduced over 3,400 active clinical stage projects; from these projects, nearly 70% are helmed by small, newer companies.[11] Compared to past cycles, the current industry pipeline has delivered on its potential, and provides an appealing source for deep investment. With over 2,500 current biotech companies within the US alone, multiple financial metrics, primarily surrounding company assets, have also allowed for the development of new dynamic investment strategies.[12] In the present state of market capitalization possible in the biotech industry, such calculated approaches can provide investors with efficient and adaptable methods of establishing both short-term and long-term portfolio growth.

## 3. Application of Potential Asset Allocation Strategies

Dynamic asset allocation refers to the integrated role of both passive and active strategies to be able to coincide with business decisions and ventures. Specifically, it involves what assets should be held, sold, and purchased to maximize financial security and reward. Financial ratios are key in these aspects, as they can combine the key data provided by the company to determine health, current financial condition, and projected prosperity. The P/E ratio utilizes the cash flow statement, balance sheet, and income statement to provide a comprehensive quantitative view of the company. When the inverse squared of this value is taken, the earnings yield may be calculated, an important valuation index directly proportional to the type of returns that a business makes.

However, decisions involving the mentioned dynamic asset allocation methodology are complex and require a deeper look in order to eliminate decisions that may negatively impact current asset holdings. Thus, knowing which stocks to buy and when to allocate them is essential in having investment success. The three main asset classes are equities, bonds, and cash equivalents, each of which allows for the asset mix to be brought back to the long term objective via dynamic allocation. This accordingly reduces risks in asset fluctuation while also providing returns much greater than the desired benchmark. But in order to optimize the different risk and return characteristics for each individual class equivalent, conservative selections must be made; adjustments to asset allocation within portfolios must also be conservative in order to retain distance from volatility, something associated with alternative equity investing. Such attention to detail and appropriate judgement is crucial for making optimal investment decisions.

In order to provide a quantitative application of asset allocation, a biotechnology company must be analyzed first by looking at the trend of assets over time, how it compares to industry standards, and its ability to easily liquidize. A sample industry in the market can be analyzed to provide a representative sample of the behavior of the market as a whole, and the efficiency of asset allocation can be assessed. Take for example the following leader in the biotechnology industry, Gilead Science Inc., and their balance sheets for the past four years. [13]

Due to the 107 billion-market cap and maturity of this biotech company, Gilead Sciences Inc. is at low risk at bankruptcy, and can maintain a low liquidity as a result. Dissecting the assets via vertical analyses is a major piece of evidence for further allocation; the decreasing liquidity and quick and current ratios may be seen as the cash and other equivalents were curbed in 2016 to about 70 percent of the previous year and as well as a decrease in inventory. Furthermore, there are small increases in other current assets and an overall 4 million decrease in current assets. Upon closer inspection, the money has been allocated to other long term assets such long-term investments, with a 20 times increase over the course of only two fiscal years (2014-2016), and also towards shareholders via dividends.

**Gilead Science, Inc.**  **Balance Sheet**
Ending December 31, 2016, December 31, 2015, December 31, 2014, December 31, 2013

| Assets | 2016 | 2015 | 2014 | 2013 |
|---|---|---|---|---|
| **Current Assets** | | | | |
| Cash and Cash Equivalents | 8,229,000 | 12,851,000 | 10,027,000 | 2,113,000 |
| Short-Term Investments | 3,666,000 | 1,756,000 | 101,000 | 19,000 |
| Net Recievables | 5,371,000 | 6,682,000 | 5,143,000 | 2,513,000 |
| Inventory | 1,587,000 | 1,955,000 | 1,386,000 | 1,697,000 |
| Other Current Assets | 1,592,000 | 1,518,000 | 1,057,000 | 655,000 |
| Total current assets | 20,445,000 | 24,762,000 | 17,714,000 | 6,997,000 |
| **Fixed (Long-Term) Assets** | | | | |
| Long-term investments | 20,485,000 | 11,601,000 | 1,598,000 | 439,000 |
| Fixed Assets and Goodwill | 4,037,000 | 3,448,000 | 2,846,000 | 1,285,000 |
| Intangible and Other Assets | 12,010,000 | 11,905,000 | 12,506,000 | 12,808,000 |
| Total fixed assets | 36,532,000 | 26,954,000 | 16,950,000 | 14,532,000 |
| **Total Assets** | 56,977,000 | 51,716,000 | 34,664,000 | 21,529,000 |

*Figure 3.* Balance sheet of Gilead Science Inc. for current and fixed assets over 4 year growth period. This statement provides a comprehensive overview of the largest individual asset components within the biotech company. By understanding the central asset distribution of Gilead, which is representative of multiple growth trends present in many biotech markets, a framework that relies most heavily on asset management can be developed.

Although it is crucial to instill trust within investors, Gilead Sciences may have to allocate money towards more of its more liquid assets and cut back on the amount of dividends as the company is diversifying to the point where growth is stagnant since too much money is given back to the investors. Furthermore, they should supplement this allocation with less equity financing and more debt financing as equity financing provides too many ownership shares into the market, which could backfire and in extreme cases, allow shareholders to effectively take hold of the company. Taking on debt financing is effective due to the low interest rates and especially the good financial standing of the market, putting it at less risk of bankruptcy. Such effective allocation can provide remedies to three major problems in biotechnology industries in general: a decrease in company growth, an abundance of competing biotechnology, and an increase in sentiments towards medical costs for American citizens. The repercussions of the aforementioned asset allocation strategies may be applied to each of these problems.

Firstly, asset allocation can assist in reversing the stagnant growth or deterioration of a company. For example, Gilead Sciences has been increasing in total assets at a rapid rate; however, the allocation towards more long term and intangible assets has had a heavy toll on the company in the previous two years as they have been stockpiling on their current liabilities, with decreasing current assets to match them.[14] What this can result in is a liquidity crisis, in addition to further financing through debt or equity financing to pay off the liabilities. This may induce the negative consequences of both types of financing to a great extent. Excessive loans may have to be payed constantly if debt financing is chosen, and with a higher interest rate, it can be burdensome for biotechnology companies. Through equity financing, more company shares may be in the biotechnology market to the point where the increasing amount of investors has too much of a hold on the company, and the diversification shares too much of the company's profit.

The biotechnology market is expanding given growing innovation. As a result, it has engendered the emergence of other large companies in the market and many startups, changing the biotechnology market from a slight oligarchy to a competitive one.[15] Accordingly, biotechnology giants like Gilead Sciences Inc. and Amgen Inc. are gaining competition to rival their products, decreasing their sales. The minimal price control they exerted is then eliminated, forcing them to decrease them to have sales for sufficient profit. This is reflected quantitatively through the decrease in gross profit of Gilead Sciences Inc. on its annual income sheet by 2 million dollars in the past year, compared to continued increases in the years prior. Asset allocation for stockpiling on more liquid assets is effective due to the decreasing income. With not as much income generated as before by major biotechnology companies, much has to be allocated to cash and accounts receivable; this is attributed to the fact that the liabilities are still increasing at a constant rate. If such allocation does not take place, a liquidity crisis could cause the company to withdraw or convert long-term bonds and fixed assets to quick cash - an expensive and unfeasible process.

Finally, the necessity of a strategy change in the biotechnology industry is reflected through increasing sentiments against biotech markets via fundamental and sentiment analysis. With the skepticism of new drugs and GMOs, the public has been induced to stray from such products, harming the sales of biotechnology companies.[16]

In curbing long term repercussions of the growing biotechnology industry, assisting in maximizing assets and revenue, and assisting long-term company growth, asset allocation has been an efficient method. Not only will the upwards trend in this industry be beneficial for the company owners, but also for the investors of the company, as it will instill trust within them (primarily through the increased dividends provided from company growth, increasing earnings per share). Although asset allocation is effective, pairing it with class diversification strategies make it an a even more dynamic methodology, which will be presented in the following sections.

## 4. Application of Potential Class Diversification Strategies

Asset class diversification provides investors with the ability to reduce overall investment risk and avoid asset damage by mixing a variety of investments within a portfolio. By doing so, investors are able to effectively control risk management of their portfolios through calculated allocation, increase the possibility of a greater profit margin, and minimize the possibility of widespread loss. There are two primary methods of asset class diversification: horizontal and vertical. Horizontal diversification includes investing in assets of similar types or classification, while vertical diversification involves investing in assets of varied classes. Within the entire range of more specialized diversification strategies, however, the asset class is a characteristic that is widely overlooked. Investors have accordingly understood that singular selection of asset classes within a portfolio is much more important than identifying individual assets or even market timing.[17] The dynamic freedom specific class selection provides is a viable strategy that can allow for a broader view for any investment framework.

The safest and most reliable approach to implementing class diversification strategies is to distribute assets broadly within a portfolio. Effective diversification requires deep assortment across multiple classes, including large-cap and small-cap stocks, corporate and government bonds, and international and emerging market stocks. However, selective diversification must be made to identify the most viable and least volatile subsector for each given market. In the case of the biotechnology industry, in which new, highly volatile products suggest short-term large profits and losses, the role of a specific asset must be understood prior to investment. For example, investors looking to maximize profits, high-yield bonds are most suitable in complementing stationary holdings, despite their increased risk. Additionally, despite some general variation in weak and strong performances, commodities and their dissociation from certain classes allow for them to serve as useful class portfolio diversifiers. In conjunction with one another, deeper approaches to wide class diversification increases potential for better long-term performance whilst retaining minimal risk.

The implementation of class diversification extends from the previous individual asset allocation, but rather than traditional short term assets, it diversifies the use of other financial assets in the market such as stocks, bonds, and commodities (the main classes). In the case of a growing biotechnology market, diversification of stocks is crucial in dictating how a company performs in the face of a recession or expansion. We again take Gilead Sciences as the representative company of the industry and analyze its market in a 52-week period with regards to earnings per share and shares outstanding, as illustrated in Figure 4.[18]

Simple moving averages over the course of 20 and 50 days were added, as indicated by the green and black lines on the stock prices. The intersection indicates change in the direction of the stock, and although there has been a cumulative increase from $71.65 to $85.1 over the course of only a few weeks, the two moving averages are set to intersect in the near future, indicating a decrease.

This is where class diversification comes in: in order to rival the setback, Gilead Sciences, along with other biotechnology giants, must be able to invest in other leading startups and increase their investments class. Additionally, they have to be mindful of the economic effects that an integrated economy has (with the UK and Europe), increasing the frequency of burst of economic growth and recessions. In a conditional inflationary environment, commodities and cash asset classes should be allocated more carefully compared to long-term bonds, which are more beneficial in a deflationary environment. This shows that asset class diversification is not static, and rather highly dynamic, especially in new markets with increasingly diverse innovations and products that are quick to change.

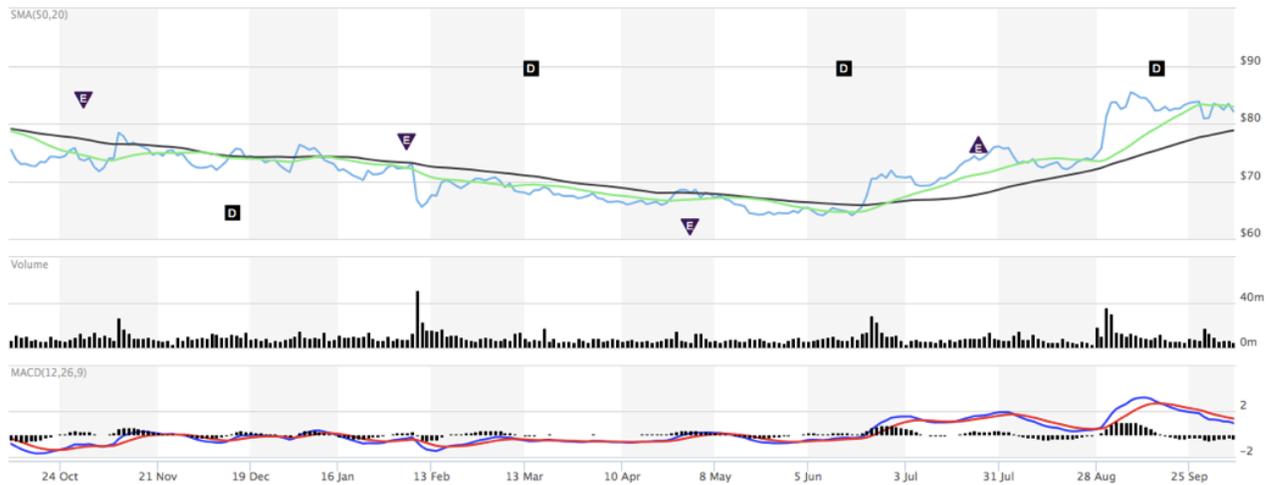

*Figure 4.* Functional market evaluations over a 52-week period beginning on October 1st for Gilead Sciences Inc., a leading biotechnology firm. Illustrated are simple moving averages over 20 and 50 days, integrated with a changing price of the stock over the course of the year.[18]

Ideal class diversification is able to accommodate for the increasing competition in the biotechnology industry and can help in increasing the amount of returns the company receives in the long term, even if there are minimal expenses from transactions in allocation. Through the allocation of stocks and bonds, the company is able to diversify further through more long-term financial assets and also retain some short-term assets in order to pay off current liquidities. Combining this with individual asset allocation, asset class diversification can assist in increasing the growth of the company, maintain a high liquidity with regard to economic recessions, and counter the recessions on the biotechnology industry from fundamental crises concerning the products and the costs to citizens.

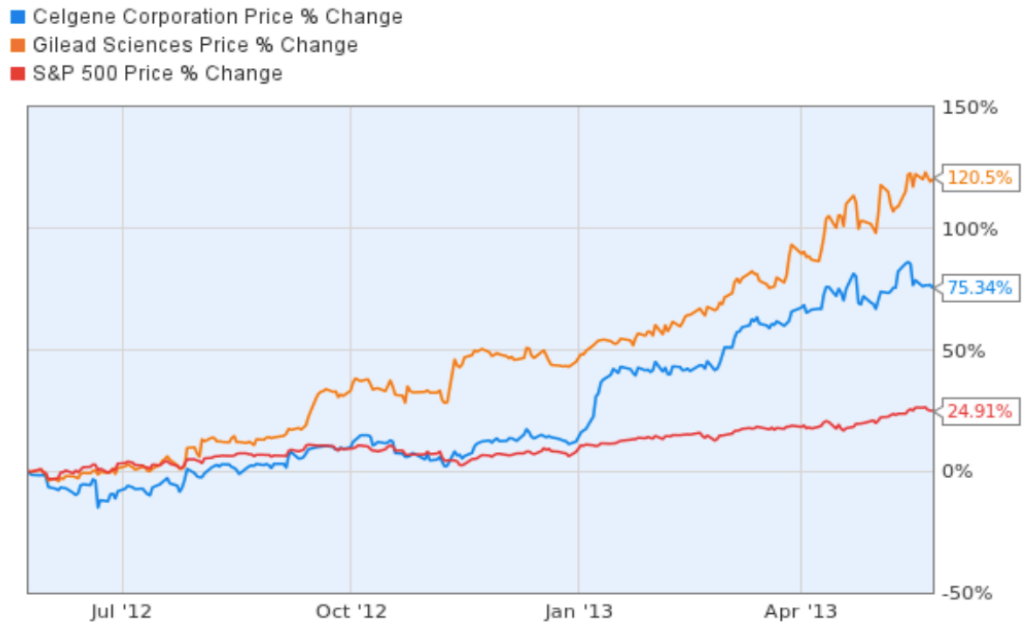

*Figure 5.* Increase in Gilead Science Inc. and other lead biotechnology firms in terms of prices as represented by percentages. This is a direct consequence of effective asset allocation and class diversification of long-term stocks.[19]

## 5. Development of a Market-Oriented Investment Framework

Although the strategies for the biotechnology industry have been identified, a framework must be developed and applied in order to identify what individual asset allocation and asset class diversification should take place. In order to determine what percentage of total assets each asset is worth, vertical analysis of the balance sheet can provide an accurate measure. In the aforementioned Gilead Sciences Inc. scenario, one may notice a heavy increase in liquid assets such as net receivables in cash during 2014 (approximately 20% increase in cash). However, the percentage of total assets in cash was cut in half over the course of two years, displaying the decreasing liquidity.

Although it is a lead company, storing away into long-term assets may not be an efficient move as a liquidity crisis may strike. Moreover, the decreasing revenue from $7,794 to $6,505 from 2015 to 2016 and corresponding decreasing in the EPS from 2.53 to 2.05 signifies decreasing trust in the company.[20] In order to instill trust within its investors and maintain a fair share of equity financing, assets must be allocated in four ways: 1) more allocation from intangible to long term investments, 2) transfer of revenue directly into cash and cash equivalents for increasing liquidity (to counter the decrease in the current and quick ratios), 3) increase in dividends to instill trusts within key investors, and 4) increase in the asset class of long-term stocks and bonds to fully utilize the time value of money in order to promote growth, which relies on the future boom of the biotechnology industry and the subsequent increasing revenue. These strategies must also be dynamically coordinated depending on external factors as well as recessions and expansions, fundamental attacks on the products of biotechnology industries from politicians, and increasing competition or entry of startups (change in the Herfindahl-Hirschmann index to determine competitiveness of the market). Effectively, combining such strategies along with the vigilance of market timing and stock directions can assist a company in devising an efficient and effective portfolio in the biotechnology industry so that they can compete with other companies.

Our optimized framework provides investors with a dynamic approach of developing their investment strategies specifically for multiple biotech markets. As it is effective in taking advantage of widespread market capitalization as exhibited throughout the biotech industry, it also retains financial stability and reduces likelihood of large loss despite market volatility. Additional strategies can be incorporated into our framework in future studies, which may allow for even further improved investment versatility. By doing so, not only would dynamic asset allocation and class diversification strategies be present, but supplemental approaches such as momentum selectivity or pairs trading would continue to improve the effectiveness and efficiency of the framework.

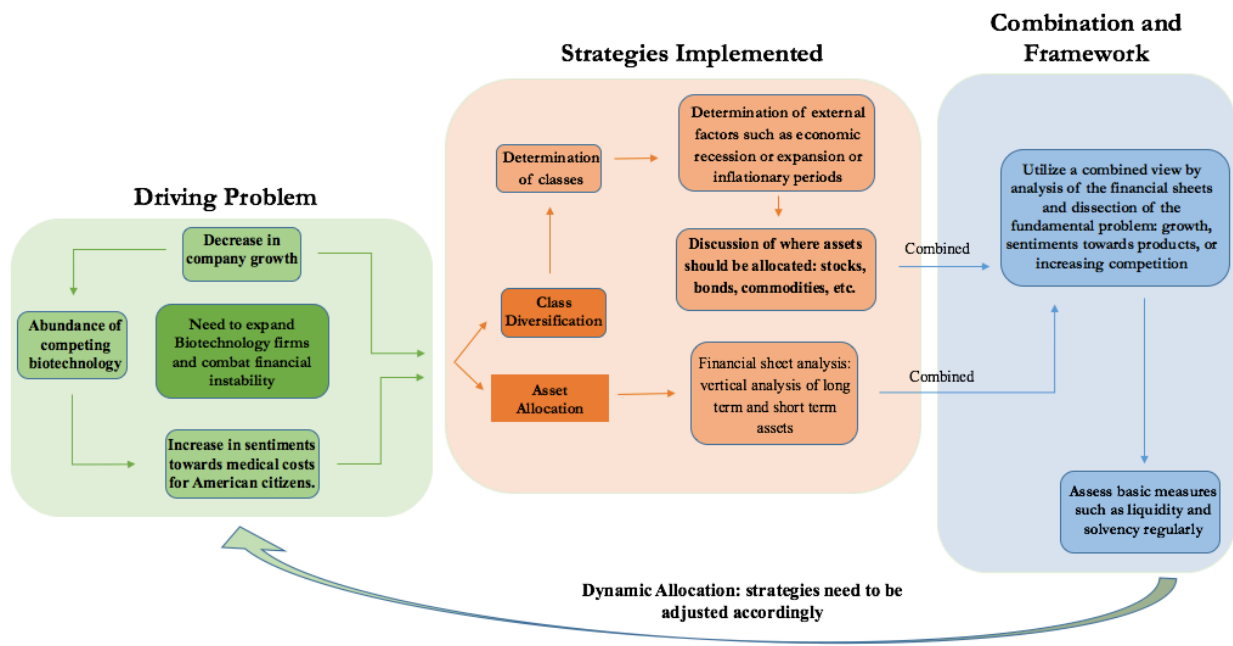

*Figure 6.* Optimized summary framework dissecting the different strategies and implementation of various asset allocations. Provides a plan for analysts to follow and apply. Initially starts with interpreting the driving problem behind the biotechnology firm under inspection and then proceeds with diversification and allocation of assets and asset classes to finally devise a solution through financial metrics analysis.